\documentclass[12pt]{article}
\usepackage{pdflscape}
\usepackage{tikz,natbib,xfrac,amsmath,amsthm,booktabs,subfigure,anysize,graphicx,float,enumitem,eurosym}
\definecolor{dark-red}{rgb}{0.4,0.15,0.15}
\definecolor{dark-blue}{rgb}{0.15,0.15,0.4}
\definecolor{medium-blue}{rgb}{0,0,0.5}
\definecolor{ChadBlue}{rgb}{.1,.1,.5}
\definecolor{ChadDarkBlue}{rgb}{.1,0,.2}
\definecolor{ChadBlue}{rgb}{.1,.1,.5}
\definecolor{ChadRoyal}{rgb}{.2,.2,.8}
\definecolor{ChadGreen}{rgb}{0,.4,0}    % Dark Green
\definecolor{ChadRed}{rgb}{.5,0,.5}  % purple
\usepackage[T1,hyphens]{url}
\usepackage[colorlinks,urlcolor=medium-blue]{hyperref}
\hypersetup{
	colorlinks, linkcolor={ChadRed},
	citecolor={ChadBlue}}
\usepackage[margin=1in]{geometry}
\usepackage{xcolor}
\usepackage{bm}

\usepackage[font=small,labelfont=bf,justification=justified,singlelinecheck=false]{caption}
\usepackage{amssymb}
\usepackage{amsfonts}
\usepackage{amsmath}
\usepackage{setspace}  %using this package conflits with footnote hyperlink jump. Using this package makes the footnote links inactive.
\usepackage[all]{hypcap}        % needed to help hyperlinks direct correctly;

\graphicspath{{figures/}}
\usepackage[english]{babel}
\usepackage{longtable}
\usepackage[singlelinecheck=off]{caption}
\usepackage[singlelinecheck=false]{caption}
\usepackage{array}
\usepackage{float}

\usepackage [autostyle, english = american]{csquotes}
\MakeOuterQuote{"}
\usepackage{chngcntr}
\usepackage{rotating, graphicx}
\usepackage{epstopdf}
\usepackage{threeparttable}
\usepackage{sectsty}
\sectionfont{\large}
\usepackage[titletoc,title]{appendix}

\newcommand{\beginappendix}{%
	\setcounter{table}{0}
	\renewcommand{\thetable}{A\arabic{table}}%
	\setcounter{figure}{0}
	\renewcommand{\thefigure}{A\arabic{figure}}%
}
\usepackage{longtable}
\usepackage{booktabs}
\usepackage{array}
\newcolumntype{R}{>{\raggedleft\arraybackslash}p{2.5cm}}
\newcolumntype{Q}{>{\raggedright\arraybackslash}p{6cm}}
\usepackage{tabu,tabularx}
\usepackage{dcolumn}
\usepackage[T1]{fontenc}
\title{Exporters' reaction to positive foreign demand shocks\thanks{I benefited enormously from conversations with Guillem Burset, Elena M. Cores, Daniel de Miguel, Mar\'{i}a Naranjo, Pedro Olivares, and Diego Rodr\'{i}guez about the swine sector. I also thank the comments and suggestions from Francisco Requena and participants at XXIII Encuentro de Econom\'{i}a Aplicada and XXII Conference on International Economics. I acknowledge the Department of Customs and Excise of the Spanish Tax Agency (AEAT) and the Spanish Chamber of Commerce for providing customs data and Aitor Garmendia for providing Spanish firms' balance sheet data. I also gratefully acknowledge the financial support from the Spanish Ministry of Science, Innovation and Universities (RTI2018-100899-B-I00, co-financed with FEDER) and the Basque Government Department of Education, Language Policy and Culture (IT885-16).}}

\author{\large {Asier Minondo}\thanks{Deusto Business School, University of Deusto, Camino de Mundaiz 50, 20012 Donostia - San Sebasti\'{a}n (Spain). Email: \href{mailto:aminondo@deusto.es}{aminondo@deusto.es}}  \\}

\date{ \today \\  }
\begin{document}
\maketitle

\begin{abstract}
I use the quasi-natural experiment of the 2018 African swine fever (ASF) outbreak in China to analyze swine exporters' reaction to a foreign market's positive demand shock. I use the universe of Spanish firms' export transactions to China and other countries, and compare the performance of swine and other exporters before and after the ASF. The ASF almost tripled Spanish swine exporters' sales to China. Swine exporters did not increase exported product portfolio or export revenue concentration in their best-performing products in China after the ASF. The increase in exports to China positively impacted export revenue and survival in third markets. This positive impact was especially intense for small swine exporters. Domestic sales also increased for swine exporters with liquidity constraints before the ASF.   
	
\end{abstract}

\begin{flushleft}
	\textbf{JEL}: F10, F14\
\end{flushleft}
\textbf{Keywords}: exporting firms, positive demand shock, China, Spain, swine.

\newpage %\setcounter{page}{1}
\onehalfspacing

\section{Introduction}
\label{sec:introduction}

%A clear context is needed to obtain a good identification

%Previous studies mix effects that may go into any direction or if in one direction (a shock may increase demand in one country and reduce it in the other, multiple effects) it is common to all goods, or use instruments that are not strong (e.g. tariffs)

% intra-firm adjustments

%There is a debate on whether a positive demand shock in a market increases or decreases a firm's sales in other markets. The standard model of international trade with heterogeneous firms predicts that Some authors find that an increase in foreign sales raises a firm's domestic sales \citep{berman2015export,erbahar2020two}. Contrarily, other scholars conclude that a decrease in domestic sales leads firms to increase their foreign sales \citep{blum2013occasional,almunia2021ventingout}.  

There is a debate on whether a positive demand shock in a market increases, decreases, or has no effect on a firm's sales in other markets. The standard model of international trade with heterogeneous firms, \cite{melitz2003impact}, assumes that markets are independent and, therefore, predicts that a positive demand shock in a market does not affect sales in other markets. However, the findings of the empirical literature are inconsistent with this prediction. Some scholars show that an increase in foreign sales raises a firm's domestic sales \citep{berman2015export,erbahar2020two}. Contrarily, other scholars conclude that a decrease in domestic sales leads firms to increase their foreign sales \citep{blum2013occasional,almunia2021ventingout}. These conflicting results indicate that the debate is far from being settled.

%To inform this debate, it is important to use a setting that enables a clear identification on how sales between different markets interplay.

%Exporters operate in an unstable environment. Tariff and non-tariff barrier changes, exchange rate shifts, transport cost variations, demand swings, and new competitors oblige exporters to regularly revise their decisions on how much to sell, how many products to offer, and what countries to serve. Firms' responses to these changes must be accurately identified because they reveal exporters' operating costs abroad, the magnitude of their market power, and the extent to which foreign shocks can spread to local markets. 

%This paper contributes to this debate using a setting that enables a clear identification on how sales between different markets interplay. Furthermore, I analyze how sales interact across foreign markets, rather than between the domestic and foreign market. 

%This paper uses a quasi-natural experiment to identify how a foreign market's unanticipated positive demand shock, namely, the 2018 African swine fever (ASF) outbreak in China, changed exporters' behavior. 

%This paper contributes to the debate using a setting that enables a clear identification on how sales between different markets interplay and analyzing how sales interact across foreign markets rather than between the domestic and the foreign market. 

%I use the 2018 African swine fever (ASF) outbreak in China as a quasi-natural experiment to identify how an unanticipated positive demand shock in a market changed exporters' sales in other foreign markets.

This paper uses a quasi-natural experiment to identify how an unanticipated and large positive demand shock in a foreign market, namely, the 2018 African swine fever (ASF) outbreak in China, changed exporters' sales in the affected market, in other foreign markets, and in the domestic market. The Chinese government ordered the slaughtering of a large share of the hog population to arrest the spread of the ASF, triggering an increase in demand for swine imports. The ASF outbreak in China affected swine products only and thus enables to employ a difference-in-differences strategy to evaluate how the ASF outbreak changed swine exporters' behavior relative to that of other exporters to China. 

Using data on the universe of Spanish firms' export transactions with China and other countries, I document that the ASF significantly increased Spanish swine exporters' export revenue in China, relative to other exporters to China. This increase stemmed from a large rise in exported quantities and moderate price growth. The phenomenon was unaccompanied by a broadening of swine exporters' product portfolio offered to China and increased export revenue concentration in firms' best-performing products. The ASF also reduced swine exporters' hazard of ceasing to export to China.

I also find that Spanish swine exporters increased their export revenue in third markets after the ASF, relative to other exporters. This increase suggests that the increase in exports to China enabled swine exporters to ease a constraint that affected exports to other destinations. Most of swine products that Spain ships to China are frozen. Due to the increased exports to China triggered by the ASF, Spanish exporters raised their freezing capacity, easing the constraint of offering frozen swine products to other markets. I show that exports of frozen swine products to third markets increased more than non-frozen swine products after the ASF. Furthermore, I find that the positive effect of the ASF on frozen swine products' exports in third markets concentrated on small swine exporters. This result is consistent with a narrative where the freezing-capacity constraint was more salient for small swine exporters, because they generated less export revenue to cover the fixed cost of investment in freezing capacity.

%I provide evidence that this constraint was related to swine exporters' freezing capacity.
%I further document a reduction in the probability that a Spanish swine exporter to China ceased exporting to third markets after the ASF outbreak.

%The reduction in the probability of exiting a third market after the ASF suggests that 

I also explore the existence of within-firm mechanisms linking foreign markets on the extensive margin. \cite{albornoz2021tariffhikes} argued that firms may incur fixed operating costs abroad, which are independent of the number of markets served by the firm. Thus, export revenue increase after the ASF is expected to cover these fixed costs and reduce the probability of ceasing to export to other foreign markets. Similarly, the positive effect of the ASF would be noticeable for small exporters because they generate less export revenue to cover the export fixed costs than large exporters. In line with this argument, I find that the ASF significantly reduced the risk of ceasing to export to third markets for small swine exporters only. 

Finally, for a limited sample of exporters to China, I analyze whether the ASF had an impact on domestic sales. I show that, on average, the ASF had no effect on swine exporters' domestic sales. However, I find that the ASF caused an increase in domestic sales for swine exporters with liquidity constraints before the ASF. This result suggests that the windfall export revenue from China provided liquidity-constrained firms with working capital to expand their domestic operations.

%This would be the case if foreign demand variations affect firms' financing conditions in the short-run, by easing their access to external finance or providing them with internal funds
%These results indicate that firms can become more competitive in markets that are not directly hit by a positive demand shock. {\color{red} Confirm that there are many variable costs that become lower?}

%{\color{red} Nice to show whether the spillover effects are larger for small firms, such as in Berman et al.. Buy hey these are large firms!!!. So our results, being based on swine exporters confirm that this cost-sharing happens also for large firms. So complementarity is not driven by small firms, such as in \cite{berman2015export} and \cite{bardaji2019exportsdomesticFrance} }

This study makes two contributions to the literature that analyzes the effect of foreign market demand shocks on exporters' behavior. First, previous studies analyzed the interplay between domestic market demand shock and foreign market sales, and between foreign market demand shock and domestic sales \citep{vannoorenberghe2012constraints,blum2013occasional,berman2015export,erbahar2020two,almunia2021ventingout}. I contribute to the literature analyzing how foreign market demand shock affects other foreign market sales.\footnote{This paper follows \cite{aranguren2021exportmarkets} who analyzed the correlation between export revenue changes in a firm's top market and the export revenue in its remainder foreign markets.} I reveal that an unanticipated and large positive demand shock in a foreign market increases export revenue in other foreign markets and reduces the risk of exiting them. These findings reveal the existence of within-firm mechanisms that transmit a foreign market shock to other foreign markets. One mechanism is related to the availability of export specific equipment. The positive demand shock from China led swine exporters to enhance their freezing capacity, which eased a constraint of selling frozen products to China and other foreign markets. Another mechanism is the existence of fixed export costs that are unrelated to the number of markets served by the firm. I provide evidence supporting the existence of these mechanisms and show that they are relevant to small exporters as expected. I also contribute to this literature providing further evidence that export revenue windfalls can alleviate firms' liquidity constraints, enabling them to expand their domestic sales.

Second, contrary to \cite{mayer2020productmix}, I find that an increase in a foreign market's demand did not lead exporters to broaden their exported products portfolios and skew their sales toward their best-performing products. This result indicates that the characteristics of goods affected by a demand shock may determine exporters' responses to a demand increase. \cite{mayer2020productmix} focused on manufactures and assumed a monopolistically competitive market structure, productivity heterogeneity across firms and products within a firm, and price elasticity of demand that decreased with quantity consumed. In this environment, firms respond to foreign demand increase extending exported product range and augmenting the revenue share of their lowest-cost products. In contrast, swine exporters sell homogeneous products that share similar marginal costs. A positive demand shock does not change export revenue distribution or product range when the increase in demand is symmetric across products.

This paper is structured as follows. Section~\ref{sec:asf_outbreak} describes the ASF outbreak in China, its impact on China's swine imports, and the evolution of Spanish swine exports to China. Section~\ref{sec:regressions} introduces the difference-in-differences regression equation and presents the results on the impact of ASF on Spanish swine exporters' behavior in China, other foreign markets, and the domestic market. Lastly, Section~\ref{sec:conclusion} concludes.

\section{African swine fever outbreak in China and response of Spanish exporters}
\label{sec:asf_outbreak}

The ASF is a highly contagious and deadly viral disease that affects domestic and wild pigs. The first ASF outbreak in China was reported on August 1, 2018, on a pig farm in Shenyang \citep{ma2020asf}. The virus spread from this city in the northern province of Liaoning to the rest of the country. According to the Food and Agricultural Organization of the United Nations (FAO), the pig population in China declined by 27\%, from 434 million in 2018 to 316 million in 2019. China notified 27 ASF outbreaks to the World Organization of Animal Health in 2020, indicating that the ASF outbreak had not been completely eradicated.\footnote{Information on the notified ASF outbreaks is available at:  \url{http://www.oie.int}}

In 2017, the year before the ASF outbreak, China had the largest stock of pigs in the world, that is, at 442 million heads, which was significantly higher than those of the next countries in the ranking: the US (73 million), Brazil (41 million), and Spain (30 million).\footnote{Data from FAO are available at: \url{http://www.fao.org/faostat/en/\#data}.} The reduced pig population due to the ASF triggered a raise  in the price of pork. According to the Chinese Ministry of Agriculture and Rural Affairs the wholesale price of pork rose by 236\%, from a minimum of 15.6 RMB/kg in May 21, 2018, to a maximum of 52.4 RMB/kg on November 1, 2019. By the end of 2020, the price of pork remained high relative to the pre-ASF period, that is, at 32.9 RMB/kg.\footnote{\url{http://www.thepigsite.com/news/2020/12/chinese-pork-prices-rise-a-little-despite-herd-recovery}}

%The stock of pigs in Spain was 30.8 million in 2018 and 31.2 in 2019. Not much of an increase.

\begin{figure}[h!]
	%\begin{center}
		\caption{\centering Trade in swine before and after the ASF, 2017-2020 \\ August 2018=100}
		\label{fig:trade}
		\begin{tabular}{c c}
			A. Imports of swine by China& B. Spanish swine exports to China \\
		\includegraphics[height=3.0in]{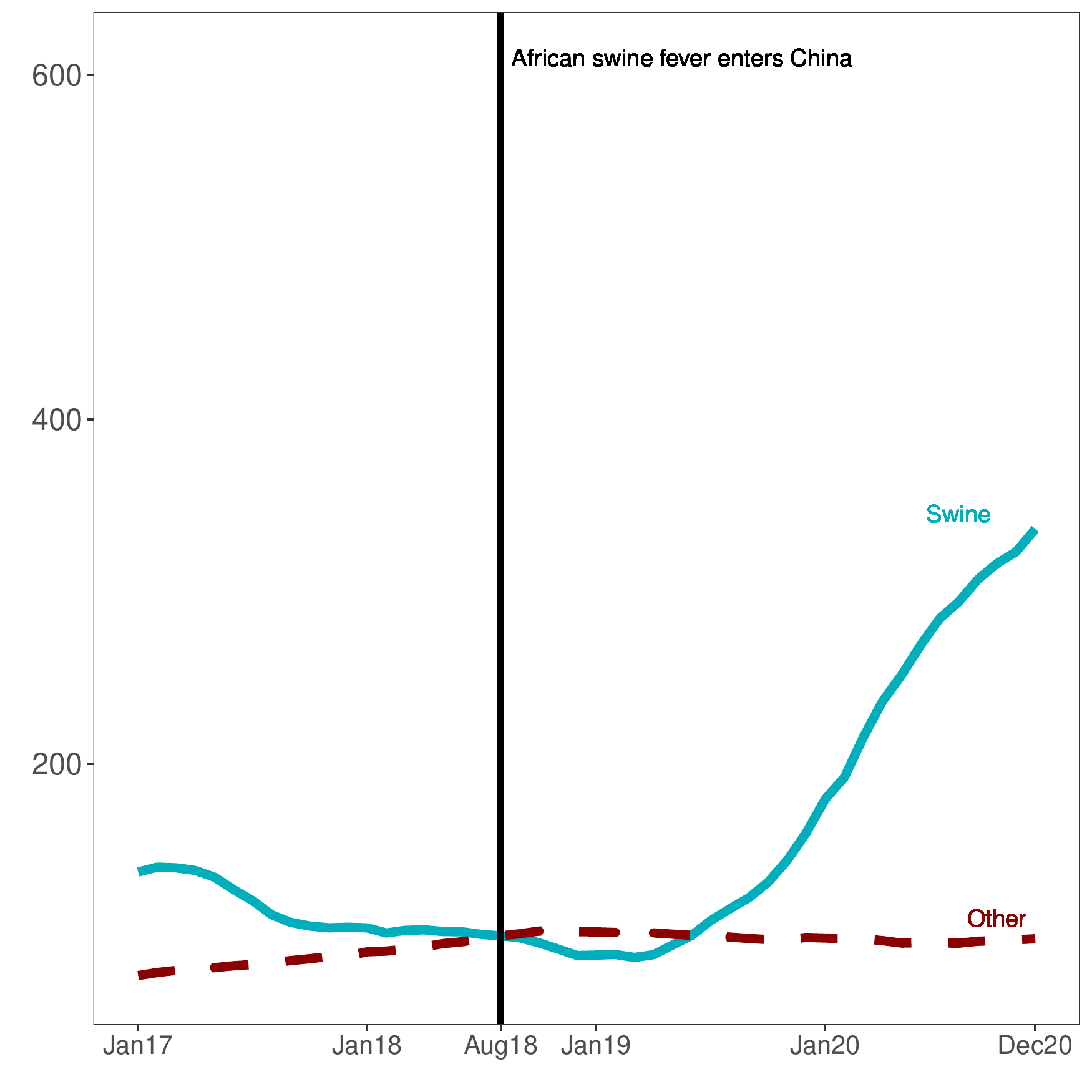}&\includegraphics[height=3.0in]{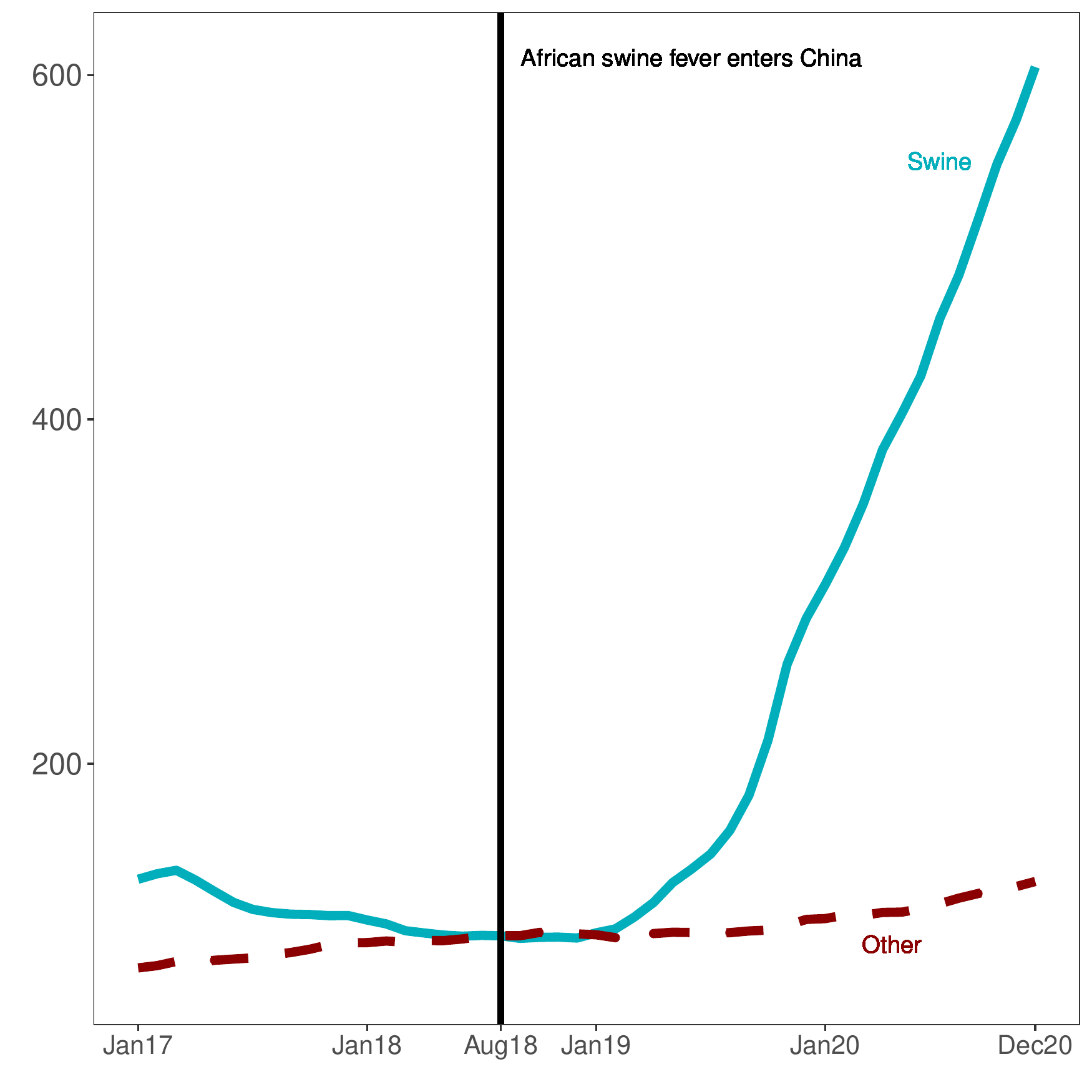}\\
		\end{tabular}
		
		\footnotesize Source: author's own elaboration based on Comtrade, China Customs, and Spanish Customs. Note: The value of trade for each month is the sum of the value of trade for that month and the previous 11 months. Swine covers all the CN 4-digit and 6-digit products included in Table~\ref{tab:swine_products}. CN 8-digit products are not included, because Comtrade and China Customs do not provide data with this level of disaggregation.
	%\end{center}
\end{figure}

The ASF also caused a substantial increase in the swine product import. Panel A of Figure~\ref{fig:trade} reports China's monthly imports of swine products (hereafter, swine) between January 2017 and December 2020. Table~\ref{tab:swine_products} in the Appendix identifies all products included in the Combined Nomenclature (CN) classification that are related to swine. A 12-month moving average of imports is calculated to avoid the noise introduced by the monthly data. For example, the value corresponding to January 2017 is the sum of the value of imports in that month and the previous 11 months. The value of imports in August 2018 is set to 100, during which the first ASF case was reported in China. China's swine imports were decreasing from the pre-ASF period to March 2019. However, between that month and December 2020, swine imports increased by more than three times. The figure indicates that the ASF provoked a change in China's trend of swine imports from 2019 onward. Contrarily, imports of other products were rising before the ASF, but began to fall since the beginning of 2019. 

Panel B of Figure~\ref{fig:trade} shows the evolution of Spanish export of swine and other products to China between 2017 and 2020. Similar to panel~A, a moving 12-month value of exports is calculated and the value of exports in August 2018 is set to 100. Spain's pre-ASF swine exports to China were decreasing, but increased by six times after the ASF. The increase in Spain's swine exports to China was much larger than in China's swine imports. As in panel~A, Spain's swine exports to China began to grow from 2019 onward, that is, a few months after the first reported ASF outbreak. Contrary to swine, other exports were mildly increasing before the ASF, stagnated between August 2018 and March 2019, and moderately increased again afterward.

Spain was China's top swine supplier in 2020, accounting for 24\% of China's total swine imports, followed by the US (16\%), Germany (12\%), and Brazil (12\%). China also became the top destination of Spain's swine exports in 2020, accounting for 47\% of Spain's total swine exports, followed by France (9\%), Japan (7\%), and Italy (6\%). Swine also became Spain's top exported product to China in 2020, accounting for 37\% of its total exports to this country.

Then, I present the characteristics of Spanish swine exporters to China before the ASF and compare them with those of the rest of Spanish exporters to China (Table~\ref{tab:swine_vs_noswine_summary}). The baseline sample is composed by firms that exported every year to China during the period 2016-2020. Data on Spanish exporters are collected from the Customs and Excise Department of the Spanish Tax Agency and include the universe of Spanish exporters of goods. The data set contains a firm identifier, export destination, the CN 8-digit classification, the value of exports, and exported quantities.\footnote{A firm is an exporter if it exports at least 1,500 euros of a given product in a given market per year.} The period 2016-2018 is defined as the pre-ASF period because the ASF did not translate into a demand shock until 2019 (panels~A and~B of Figure~\ref{fig:trade}).\footnote{A firm is a swine exporter if it exported to China any of the products included in Table~\ref{tab:swine_products} in all the pre-ASF years. A firm is an exporter of other products if it exported to China in all the pre-ASF years, but did not export any product included in Table~\ref{tab:swine_products}.} 

%Clean the table the first time I compile it
\begin{table}[tbp] \centering
\newcolumntype{C}{>{\centering\arraybackslash}X}

\caption{Swine vs. other exporters to China before the ASF}
\label{tab:swine_vs_noswine_summary}
{\small
\begin{tabularx}{\textwidth}{lrr}

\toprule
\multicolumn{1}{l}{}&\multicolumn{1}{p{2.5cm}}{\raggedleft Swine exporters}&\multicolumn{1}{p{2.6cm}}{\raggedleft Other exporters} \tabularnewline
\midrule\addlinespace[1.5ex]
Number of exporters to China&46&2,163 \tabularnewline
Average export revenue (million euros)&53&48 \tabularnewline
Average export revenue in China (million euros)&12&2 \tabularnewline
Average number of export destinations&24&29 \tabularnewline
Average number of exported products&19&20 \tabularnewline
Average number of exported products to China&7&4 \tabularnewline
Exports to China as a share of total exports (\%)&25&17 \tabularnewline
\bottomrule \addlinespace[1.5ex]
\end{tabularx}
}
\footnotesize \raggedright Note: average yearly values for the period 2016-2018. A swine exporter is a firm exporting a product included in Table~\ref{tab:swine_products} in all years between 2016 and 2020. Other exporters are firms that exported to China all the years between 2016 and 2020, but did not export any product included in Table~\ref{tab:swine_products}.

\end{table}

%\footnotesize \raggedright Note: average yearly values for the period 2016-2018. A swine exporter is a firm exporting a product included in Table~\ref{tab:swine_products} in all years between 2016 and 2020. Other exporters are firms that exported to China all the years between 2016 and 2020, but did not export any product included in Table~\ref{tab:swine_products}.

The number of swine and other (non-swine) exporters to China before the ASF was 46 and 2,163, respectively. The small number of swine exporters is partly explained by the Chinese General Administration of Customs requirement for Spanish swine slaughterhouses, processing plants, cutting plants, and cold stores to export to China.\footnote{Firms should abide by the requisites established by the protocol signed by the Spanish and Chinese governments on November 7, 2007. Spanish establishments were inspected by Chinese authorities before 2018. However, in a new protocol signed in November 2018, Chinese authorities delegated the inspection and certification processes to the Spanish Ministry of Agriculture, Fishery and Feeding.} The average yearly value of exports to all destinations by swine exporters to China during the pre-ASF period was 53 million euros, which was 10\% larger than that for other exporters. The difference between swine and other exporters was much larger regarding the average export value to China: 12 million and 2 million euros, respectively. Swine exporters to China exported to less destinations than other exporters, but the number of exported products to all destinations was similar for swine and other exporters. However, swine exporters exported more products to China than other exporters.\footnote{Exporters of swine to China also exported other products to this country. However, the average share of these products in a swine firm's total exports to China was low, that is, at 1.9\%.} Finally, China already represented an important market for Spanish swine exporters before the ASF, accounting for a quarter of their total exports.
 
%Overall, non-swine products represented only 0.5\% of swine exporters total export to China before the ASF and decreased to 0.2\% after the ASF. Finally, 

%\input{../results/summary_annex.tex}

The key identifying assumption of the difference-in-differences strategy is that other exporters provide an appropriate counterfactual of the trend that swine exports would have followed in China in the absence of the ASF outbreak. Table~\ref{tab:same_preasf_growth} presents the sales growth rate during the pre-ASF period, 2016-2018, for swine and other exporters in the three markets analyzed in this study: China, other foreign markets, and the domestic market. I test whether the pre-ASF average growth rate of sales for swine exporters is similar to the pre-ASF average growth rate of sales for other exporters. I cannot reject the null hypothesis that there is no significant difference in the average growth rate of sales between swine and other exporters in the three markets. This result indicates that the empirical results are not affected by differences in pre-trends between swine and other exporters.

\begin{table}[t!]
	\begin{center}
		\footnotesize
		\caption{\centering Common trends before ASF (export growth 2018-2016)}
		\label{tab:same_preasf_growth}
\resizebox{\textwidth}{!}{
	\begin{tabular}{lrrrrrrrrr}
	\toprule
Market&\multicolumn{3}{c}{Swine exporters}&\multicolumn{3}{c}{Other exporters}&\multicolumn{3}{c}{Difference}\\
\multicolumn{10}{c}{ }\\
&Mean&Std.Dev.&Firms&Mean&Std.Dev.&Firms&Diff&SE&$\rho$-value\\	
\cmidrule(lr){2-4} \cmidrule(lr){5-7} \cmidrule(lr){8-10}
China&4.47&15.66&46&4.27&33.93&2,163&0.20&2.42&0.93\\
Other foreign markets&0.84& 2.43&43&0.82&8.01&2,027&-0.02&0.41&0.97\\
Domestic&0.11&0.34&15&0.30&2.02&495&0.19&0.13&0.13\\
\bottomrule
\end{tabular}
}
		\caption*{\begin{footnotesize}Note: A swine exporter is a firm exporting a product included in Table~\ref{tab:swine_products} in all years between 2016 and 2020. Other exporters are firms that exported to China all the years between 2016 and 2020, but did not export any product included in Table~\ref{tab:swine_products}. Columns~(2) and~(5) report the growth rate in sales between 2018 and 2016; columns~(3) and~(6) the standard deviation, and columns~(4) and~(7) the number of firms in each group and market. Column~(8) reports the difference between the means, column~(9) the standard error the means difference, and column~(10) the $\phi$-value of the t-test.
		\end{footnotesize}}
	\end{center}
\end{table}

\section{Impact of the African swine fever on the behavior of Spanish exporters to China}
\label{sec:regressions}

I divide this section in four parts. First, I present the specification used in the empirical analyses. Second, I analyze how the ASF outbreak affected the behavior of exporters in China. Third, I explore whether the positive demand shock in China altered the behavior of exporters in other foreign markets. Finally, I study the effect of the ASF outbreak on domestic sales.

\subsection{Specification}

I use the following specification to explore whether the ASF altered the behavior of Spanish swine exporters to China:

\begin{equation}
\label{eq:did}
y_{ft}=\beta(ShareSwine_{f}*PostASF_{t})+\gamma_{f}+\gamma_{t}+\epsilon_{ft}
\end{equation}

where $y_{ft}$ is the dependent variable for firm $f$ in year $t$. $ShareSwine_{f}$ is the share of swine exports in firm $f$'s total exports to China before the ASF. It captures the extent to which an exporter was treated by the positive demand shock triggered by the ASF.\footnote{I also defined an alternative indicator variable that turned one if swine products represented 90\% or more of a firm's exports to China before the ASF. Main conclusions were not altered.} $PostASF$ turns one if $t$ is 2019 or 2020, $\gamma_{f}$ and $\gamma_{t}$ are firm and year fixed effects, respectively, and $\epsilon_{ft}$ is the disturbance term. The key variable is the difference-in-differences interaction $ShareSwine_{f}*PostASF_{t}$, which captures whether the dependent variable changed for swine exporters after the ASF. The construction of the interaction variable reflects the expectation of large change in the dependent variable after the ASF for firms whose swine products accounted for a large share in their exports to China before the ASF.

\subsection{Impact of the African swine fever on exporters' behavior in China}

%I define the beginning of the post-ASF period in 2019 and cluster standard errors at the firm level. 
%\footnote{Results are qualitatively similar if I exclude 2018 data from the sample.} 

This subsection analyzes whether the ASF outbreak altered the behavior of swine exporters in China. Panel~A of Table~\ref{tab:china} shows that swine firms' exports significantly increased after the ASF, relative to other exporters to China.\footnote{I cluster standard errors at the firm level.} For a firm whose swine exports represented 100\% of its total exports to China before the ASF, the fever outbreak led to a 175\% increase in the value of exports (exp(1.011)-1). Panels~B and~C show whether the increase in export revenue was caused by an increase in quantity or price. I calculate export price as the ratio between export revenue and exported kilograms. 86\% and 14\% of the increase are attributed to a rise in quantities and prices, respectively. These figures demonstrate that Spanish swine exporters substantially increased the amount of swine exported to China to meet the demand generated by the ASF. However, this analysis is taken with caution, because revenues and quantities of different swine products are added at the firm level.   

%In addition to the continuous \textit{Share Swine} variable, I define an alternative variable that turns one if swine products represented 90\% or more of a firm's exports to China before the ASF. This percentage is the average share of swine products in swine exporters' exports to China before the ASF. Column~2 of panel A shows that the ASF led to a 226\% increase in export revenue for firms whose swine products represented 90\% of their total exports to China before the ASF (exp(1.183)-1). 

\begin{table}[t!]
	\begin{center}
		\footnotesize
		\caption{\centering Impact of ASF on swine exporters' behavior in China}
		\label{tab:china}
		{
\def\sym#1{\ifmmode^{#1}\else\(^{#1}\)\fi}
\begin{tabular}{lcc}
\toprule
Dependent variable&Post X Share swine&Observations\\
\cmidrule(lr){1-1} \cmidrule(lr){2-2} \cmidrule(lr){3-3}
A. Revenue (log)&       1.011\sym{a}&11,045\\
&     (0.156) &      \\
[1em]
B. Quantity (log)&0.871\sym{a}&11,045\\
&     (0.165)   &    \\
[1em]
C. Price (log)&       0.140& 11.045       \\
&     (0.090)&       \\
[1em]
D. Number of exported products &       0.073 &11,045      \\
&     (0.067) &      \\
[1em]
E. Revenue concentration  &       0.017 &11,045       \\
&     (0.032) &       \\
[1em]
F. Exit probability&      -0.151\sym{a}&29,956\\
&     (0.043) &      \\
\bottomrule
\end{tabular}
}

		\caption*{\begin{footnotesize}Note: All estimations include firm and year fixed effects. Standard errors clustered at the firm level are in parentheses. a, b, and c denote statistical significance at the 1\%, 5\%, and 10\% levels, respectively. 				
		\end{footnotesize}}
	\end{center}
\end{table}

Next I explore the effect of the ASF on the extensive margin of trade. \cite{mayer2020productmix} contended that exporters would increase their exported product portfolio and skew their sales toward their best-performing product in response to a positive foreign market demand shock. I test these hypotheses in panels~D and~E. I define a product as an 8-digit code included in the CN classification. I estimate Equation~\eqref{eq:did} with the number of exported products as the dependent variable and a Poisson model.\footnote{I use the Stata command ppmlhdfe developed by \cite{correia2019ppmlhdfe}.} Panel~D shows that the number of products exported by swine exporters to China did not significantly increase after the ASF.\footnote{When computing the number of products I also include any non-swine products exported by a swine exporter.} Panel~E explores whether the positive demand shock led Spanish swine exporters to China to concentrate their export revenue on their best-performing products. Following \cite{mayer2020productmix}, I use a Theil index to measure the skewness of export revenue by product as follows:

\begin{equation}
	\label{eq:theil}
	T_{ft}=\frac{1}{N_{ft}}\sum_{p}\frac{x_{fpt}}{\overline{x_{ft}}}log\Big(\frac{x_{fpt}}{\overline{x_{f}}}\Big)
\end{equation}

where $N_{ft}$ is the number of products that firm $f$ exports to China in year $t$ and $p$ stands for an 8-digit CN product. $\overline{x_{f}}$ is defined as:

\begin{equation}
	\label{eq:theil_average}
	\overline{x_{ft}}=\frac{\sum_{p}x_{fpt}}{N_{ft}}
\end{equation}

Results reported in panel~E show that export revenue did not concentrate more on the best-performing products after the ASF. Therefore, contrary to \cite{mayer2020productmix}, I find no change in the product portfolio and export revenue distribution after a foreign market demand shock. 

The ASF's positive effect on export revenue and its absence of influence on the number of exported products, and the skewness of sales are consistent with an environment where firms sell homogeneous products, have similar costs across products, and experience a demand increase that is symmetric across products. First, slaughtered hogs are homogeneous products whose prices are referenced to auctions in wholesale markets.\footnote{For example, prices in Spain are referenced to weekly auctions at Mercolleida. \cite{rauch1999networks} also classifies swine as a homogeneous product because it is traded on organized exchanges.} Second, marginal costs are similar across products because producers simultaneously obtain different swine products (e.g., carcasses, hams, shoulders, and offal) at the slaughtering and cutting processes. Third, the range of swine products obtained after slaughtering and cutting is the same for all producers. Finally, the increase in demand from China after the ASF was fairly symmetric across the products that swine exporters were selling to China before the ASF.\footnote{The correlation between the rankings of the exported swine products before and after the ASF was 0.85.} In this context, an increase in demand for swine products will raise prices, triggering an increase in slaughtered hogs. Differences in profit rates between products will not affect export revenue distribution and exported product range because of similar marginal costs. These variables will be governed only by the product specific increase in demand from China. If the demand rises symmetrically, then revenue distribution or product range will not change.   

Finally, panel~C examines whether the swine exporters' probability of exiting the Chinese market decreased after the ASF. I estimate Equation~\eqref{eq:did} with a linear probability model with a dependent variable that turns 1 if the firm did not export to China in \textit{t}. I broaden the sample to include any firm that exported to China, at least, one year during the pre-ASF period. I find that swine exporters reduced the probability of exiting the Chinese market after the ASF. For a firm whose swine exports accounted for 100\% of its total exports to China before the ASF, the probability of exiting the Chinese market decreased by 15 percentage points after the ASF. 

%This amount represents a 25\% decrease relative to the exit probability baseline during the post-ASF period (8\%).

%\footnote{As explained in detail in Section~\ref{sec:asf_outbreak}, non-swine products represent a small share of swine firms' export revenue.}

%As explained in the Introduction section, these results are consistent with a model where firms export homogeneous products, marginal costs are similar across products, and the foreign market demand shock is symmetric across products. 

\subsection{Impact of the African swine fever on exporters' behavior in other foreign markets}
This section presents an analysis of whether the ASF outbreak in China altered the behavior of Spanish swine exporters in other non-Chinese foreign markets. Accordingly, I add-up the value and quantity of exports and the number of exported products in non-Chinese destinations of the Spanish exporters to China. I select the firms that exported every year to China and other non-Chinese markets during the period 2016-2020. 

Panel~A of Table~\ref{tab:third_markets} shows that the ASF positively impacted the export revenue in other foreign markets of Spanish swine exporters to China, relative to other exporters to China. A firm whose swine exports represented 100\% of its total exports to China before the ASF increased its value of exports to other foreign markets by 19\%. The increase in export revenue was lower than that in China, which is 175\%. Panels~B and~C show that the increase in export revenue can be attributed more to a raise in quantity than prices; however, none of the coefficients are statistically significant. Panel~D shows that the ASF reduced the number of products exported to third markets. However, the concentration of export revenue across products did not change after the ASF (panel~E). 

\begin{table}[htbp]
	\begin{center}
		\footnotesize
		\caption{African swine fever impact on other foreign markets}
		\label{tab:third_markets}
		{
\def\sym#1{\ifmmode^{#1}\else\(^{#1}\)\fi}
\begin{tabular}{lcc}
\toprule
Dependent variable&Post X Share swine&Observations\\
\cmidrule(lr){1-1} \cmidrule(lr){2-2} \cmidrule(lr){3-3}
A. Revenue (log)&       0.174\sym{b}&10,350\\
&     (0.079) &      \\
[1em]
B. Quantity (log)&0.110&10,350\\
&     (0.089)   &    \\
[1em]
C. Price (log)&       0.064& 10,350       \\
&     (0.052)&       \\
[1em]
D. Number of exported products &        -0.158\sym{a} &10,350      \\
&     (0.055) &      \\
[1em]
E. Revenue concentration  &       -0.009 &10,350       \\
&     (0.057) &       \\
\bottomrule
\end{tabular}
}

		\caption*{\begin{footnotesize}Note: All regressions include firm and year fixed effects. Standard errors clustered at the firm level are in parentheses. a, b, and c denote statistical significance at the 1\%, 5\%, and 10\% levels, respectively. 				
		\end{footnotesize}}
	\end{center}
\end{table}

The regression results demonstrate that increased export revenue from China raised swine exporters' revenue in other foreign markets. Export revenue across foreign markets can be correlated if higher sales in a foreign market ease a constraint to expand exports in other foreign markets. For swine exporters, the increase in exports to China could have eased the freezing-capacity constraint.

A total of 92\% of Spain's exported swine products to China in 2020 were frozen products. The increase in exports due to the ASF led swine exporters to enhance their freezing capacity  \citep{burset2021congelacion}. I explore whether the increase in freezing capacity eased the constraints to selling frozen products in other foreign markets. I separate firms' export revenue from swine products in two categories: frozen and non-frozen swine products.\footnote{I look to the product description in the Combined Nomenclature classification to classify swine products between frozen and non-frozen.} A comparison of column~1 with column~2 of Table~\ref{tab:frozen} shows that the ASF significantly increased only frozen products' export revenue in other foreign markets. 

%This result is consistent with the argument that the increase in freezing capacity allowed Spanish swine exporters to offer additional frozen products in third markets. 

%For swine exporters, additional resources due to the ASF may had facilitated the investment in specific export equipment, namely, freezing tunnels. 

%Frozen products accounted for 43% and 38% of Spanish exports to third markets before and after the ASF, respectively.

\begin{table}[t!]
	\begin{center}
		\footnotesize
		\caption{African swine fever impact on other foreign markets by product type and exporter size}
		\label{tab:frozen}
		{
%\resizebox{\textwidth}{!}{
\def\sym#1{\ifmmode^{#1}\else\(^{#1}\)\fi}
\begin{tabular}{lcccc}
\toprule
Products included in swine exports&Frozen&Non-frozen&Frozen&Non-frozen\\
\cmidrule(lr){2-2} \cmidrule(lr){3-3} \cmidrule(lr){4-4} \cmidrule(lr){5-5}

                    &\multicolumn{1}{c}{(1)}       &\multicolumn{1}{c}{(2)}       &\multicolumn{1}{c}{(3)}       &\multicolumn{1}{c}{(4)}       \\
\midrule
Post X Share swine  &       0.196\sym{a}&       0.164       &       0.090\sym{c}&       0.164       \\
&     (0.070)       &     (0.122)       &     (0.054)       &     (0.124)       \\
[1em]
Post X Share swine X Small exporter&                   &                   &       0.405\sym{b}&       0.001       \\
&                   &                   &     (0.176)       &     (0.306)       \\
\midrule
Observations        &       10,300       &       10,345       &       10,300       &       10,349       \\
\bottomrule
\end{tabular}
%}
}
		\caption*{\begin{footnotesize}Note: The dependent variable is the log value of exports to other foreign markets. Frozen and non-frozen swine products are identified using the product description in the Combined Nomenclature classification. All regressions include firm and year fixed effects. Standard errors clustered at the firm level are in parentheses. a, b, and c denote statistical significance at the 1\%, 5\%, and 10\%, respectively. 				
		\end{footnotesize}}
	\end{center}
\end{table}

Swine firms will invest in freezing equipment as long as the export revenue from frozen products cover the cost of investment. It is reasonable to assume that the freezing equipment constraint was greater for small swine exporters than for large ones because they generated less export revenue to cover the fixed investment cost. Therefore, the positive effect of the ASF on third markets should be especially intense for small exporters. To test this hypothesis, I define a new variable, $Small_{f}$, which turns one if the swine exporter $f$ had a total export revenue below the 25th percentile of export revenue of swine exporters before the ASF. I expand Equation~\eqref{eq:did} including a new interaction term as follows:

\begin{equation}
	\label{eq:small}
	\begin{split}
		y_{ft}=\beta_{1}(ShareSwine_{f}*PostASF_{t})+\beta_{2}(ShareSwine_{f}*PostASF_{t}*Small_{f})+  \\+\gamma_{f}+\gamma_{t}+\epsilon_{ft}
	\end{split}
\end{equation}

The triple interaction term, $ShareSwine_{f}*PostASF_{t}*Small_{f}$, captures whether the effect of ASF on swine exporters was different for small and large exporters. 

Column~3 of Table~\ref{tab:frozen} shows that the positive effect of the ASF on frozen product export revenue in third markets was much larger for small exporters than for large ones. For a large exporter, whose swine exports represented 100\% of its total exports to China, frozen swine exports to third markets increased by 9\% after the ASF (exp(0.090)-1). If the exporter was small, then the ASF would had led to a 64\% increase in exports (exp(0.090+0.405)-1). There are no differences between large and small swine exporters regarding non-frozen products.

I also analyze whether foreign markets are correlated in the probability of exit. I extend the sample to include all firms that exported to China and another foreign market at least one year during the pre-ASF period. Column~1 of Table~\ref{tab:exit_other} shows that difference-in-differences interaction coefficient is negative but statistically insignificant. Therefore, we find that the ASF outbreak in China had no significant effect on the risk of exiting other foreign markets.

\begin{table}[htbp]
	\begin{center}
		\footnotesize
		\caption{African swine fever impact on the probability of ceasing to export to other foreign markets}
		\label{tab:exit_other}
		{
\def\sym#1{\ifmmode^{#1}\else\(^{#1}\)\fi}
\begin{tabular}{l*{2}{c}}
\toprule
                    &\multicolumn{1}{c}{(1)}       &\multicolumn{1}{c}{(2)}       \\
\midrule
Post X Share swine  &      -0.021       &      -0.012       \\
                    &     (0.017)       &     (0.021)       \\
[1em]
Post X Share swine x Small exporter&                   &      -0.045\sym{b}\\
                    &                   &     (0.021)       \\
\midrule
Observations        &       37,044       &       37,044       \\
\bottomrule
\end{tabular}
}

		\caption*{\begin{footnotesize}Note: The dependent variable is the probability that a firm ceases to export to a non-Chinese market. All regressions include firm and year fixed effects. Standard errors clustered at the firm level are in parentheses. a, b, and c denote statistical significance at the 1\%, 5\%, and 10\% levels, respectively. 				
		\end{footnotesize}}
	\end{center}
\end{table}

\cite{albornoz2021tariffhikes} argued that the probability of exporting to two different destinations can be correlated if fixed export costs that are independent of the number of markets served by the firm exist. If export revenue rises in a market, then the probability of covering the fixed export costs will increase, whereas that of ceasing to export in the remaining destinations will decrease. They also predict that the effect of a rise in export revenue should benefit small exporters because they generate less export revenue to cover fixed export costs than large firms. Column~2 of Table~\ref{tab:exit_other} tests this hypothesis. As predicted by \cite{albornoz2021tariffhikes}, it shows that the ASF significantly reduced the risk of ceasing an export activity in other foreign markets for small swine exporters only.

\subsection{Impact of the African swine fever on exporters' domestic sales}

This section analyses the impact of the ASF on swine exporters' domestic sales. I calculate domestic sales as the difference between total and export revenue. Firms' total revenue is obtained from the SABI database and is matched with Customs data using the correspondence explained in \cite{delucio2018prices}. The new sample has a lower number of observations than the baseline sample due to two reasons.\footnote{The number of firms included in the sample is reduced by 77\%. These firms accounted for 36\% of the pre-ASF exports of the baseline sample.} First, the correspondence between SABI and Customs is only available for some exporters. Second, total revenue is reported by SABI with a lag. The latest data corresponds to 2019, which covers only one year of the post-ASF period.

\begin{table}[htbp]
	\begin{center}
		\footnotesize
		\caption{African swine fever impact on domestic sales}
		\label{tab:domestic}
		{
\def\sym#1{\ifmmode^{#1}\else\(^{#1}\)\fi}
\begin{tabular}{lcc}
\toprule
                    &\multicolumn{1}{c}{(1)}       &\multicolumn{1}{c}{(2)}       \\
\midrule
Post X Share swine  &       0.070       &      -0.000       \\
&     (0.048)       &     (0.034)       \\
[1em]
Post X Share swine X High-liquidity constraint&                   &       0.210\sym{a}\\
&                   &     (0.077)       \\
\midrule
Observations        &        2,040       &        2,040       \\
\toprule
\end{tabular}
}

		\caption*{\begin{footnotesize}Note: The dependent variable is the (log) value of domestic sales. All regressions include firm and year fixed effects. Standard errors clustered at the firm level are in parentheses. a, b, and c denote statistical significance at the 1\%, 5\%, and 10\% levels, respectively. 				
		\end{footnotesize}}
	\end{center}
\end{table}

Table~\ref{tab:domestic} reports the estimates. The Post X Share swine coefficient in column~1 is positive but statistically insignificant. This indicates that the positive demand shock due to the ASF had no effect on swine exporters' to China domestic sales, relative to other exporters to China. I also explore whether the impact of the ASF on domestic sales was heterogeneous across firms. Following \cite{berman2015export}, I analyze whether the windfall export revenue from China benefited the expansion of domestic sales for firms with liquidity constraints. To proxy a firm's liquidity constraint, I calculate the difference between current liabilities and current assets, and divide it by current assets.\footnote{This data is obtained from the SABI database. I calculate the average liquidity constraint during the pre-ASF period (2016-2018).} I create a dummy variable, High-liquidity constraint, that turns one if the swine exporter to China had a liquidity constraint above the median during the pre-ASF period. Column~2 shows that the ASF had no impact on domestic sales for swine exporters without liquidity constraints. However, domestic sales increased significantly after the ASF for firms that had liquidity constraints. Specifically, for a firm whose swine exports represented 100\% of its total exports to China, domestic sales increased by 23\% after the ASF outbreak.

%First, I analyze whether the effect of the ASF on domestic sales was different for small and large firms. I define an indicator variable, $Small firm$, that turns one if a firm's average total revenue before the ASF was equal of below the 25\% percentile of total revenue. Column~2 shows that the effect of the ASF on domestic sales was larger for small than large firms. However, the $Post X Share swine X Small$ coefficient is statistically not significant. 

%Second, I analyze whether the impact of the ASF on domestic sales was larger for firms whose exports to China represented a large share of their total revenue before the ASF. I calculate the ratio exports to China before the ASF/total revenue before the ASF. I defined a new indicator variable, $High Share China$, that turned one if the ratio was equal or larger to the 75th percentile of the ratio. Column~3 shows that domestic sales increased significantly after the ASF for swine firms whose exports to China represented a high share of their total revenue.

\section{Conclusion}
\label{sec:conclusion}
This paper shows that Spanish swine exporters reacted to the unanticipated positive demand shock from China by increasing heavily their exports to that market. This increase was unaccompanied by an expansion of exported product range or increased export revenue concentration in the best-performing products. Moreover, the increase in exports to China caused a raise in export revenue in other foreign markets. I argue that increased sales to China led firms to expand their freezing capacity, easing the constraints to offer frozen swine products in other foreign markets. Consistent with this hypothesis, I observe that exports in third markets after the ASF increased for frozen swine products only. Furthermore, I demonstrate that the increase in frozen exports was intense for small exporters.

I also find evidence for the existence of within-firm mechanisms linking export markets on the extensive margin. The ASF reduced small exporters' risk of ceasing to export to third markets. This reduction was especially intense for small exporters. This result is consistent with the existence of fixed export costs that are independent of the number of markets served by a firm. Finally, I show that the ASF had a positive impact on domestic sales for swine exporters with liquidity constraints.

%\clearpage

\appendix \label{app:all}

\setcounter{equation}{0}
\renewcommand\theequation{A.\arabic{equation}}

%\pagenumbering{arabic}

\setcounter{figure}{0}
\renewcommand\thefigure{A.\arabic{figure}}

\setcounter{table}{0}
\renewcommand\thetable{A.\arabic{table}}

%\clearpage

\beginappendix
\begin{appendices}

\begin{table}
\begin{center}
\caption{Swine products}
\label{tab:swine_products}
\begin{tabularx}{\linewidth}{ l X }
	\toprule
	CN Code&Description\\
	\midrule
0103&Live swine\\
0203&	Meat of swine, fresh, chilled or frozen\\
020630&Edible offal of swine, fresh or chilled\\
020641&Edible offal of swine, frozen livers\\
020649&Edible offal of swine, frozen others\\
0209&Pig fat, free of lean meat, and poultry fat, not rendered or otherwise extracted, fresh, chilled, frozen, salted, in brine, dried or smoked\\
021011&Meat of swine hams, shoulders and cuts thereof, with bone in salted, in brine, dried or smoked\\
021012&Meat of swine bellies (streaky) and cuts thereof in salted, in brine, dried or smoked\\
021019&Other meat of swine in salted, in brine, dried or smoked\\
02109941&Edible flours and meals of meat or meat offal-Livers of domestic swine\\
02109949&Edible flours and meals of meat or meat offal-Others of domestic swine\\
0502& Pigs', hogs' or boars' bristles and hair; badger hair and other brush making hair; waste of such bristles or hair\\
0504& Guts, bladders and stomachs of animals (other than fish), whole and pieces
thereof, fresh, chilled, frozen, salted, in brine, dried or smoked\\
0506&Bones\\
1501& Pig fat (including lard) and poultry fat\\
160241 &Other prepared or preserved meat, meat offal or blood. Hams and cuts thereof of swine\\
160242 &Other prepared or preserved meat, meat offal or blood. Shoulders and cuts thereof of swine\\
160249 &Other prepared or preserved meat, meat offal or blood. Other, including mixtures of swine\\
16029051&Preparations of blood containing meat or meat offal of domestic swine\\
41033000&Raw hides and skins of swine\\
41063100&Tanned or crust hides and skins of swine in the wet state (including wet-blue)\\
41063200&Tanned or crust hides and skins of swine in the dry state (crust)\\
41132000&Leather further prepared after tanning or crusting of swine\\
	\bottomrule
\end{tabularx}
\end{center}	
\end{table}

%\bigskip
%\bigskip

\end{appendices}
\end{document}